\documentclass[conference]{IEEEtran}
\IEEEoverridecommandlockouts
\usepackage{amsmath,amssymb,amsfonts}
\usepackage{algorithmic}
\usepackage{graphicx}
\usepackage{textcomp}
\usepackage{xcolor}
\def\BibTeX{{\rm B\kern-.05em{\sc i\kern-.025em b}\kern-.08em
T\kern-.1667em\lower.7ex\hbox{E}\kern-.125emX}}

\usepackage[numbers]{natbib}
\usepackage[inline]{enumitem}
\usepackage{url}

\definecolor{darkgreen}{rgb}{0,0.6,0}

\begin{document}

\title{Research Software Sustainability and Citation}

\author{\IEEEauthorblockN{Stephan Druskat}\thanks{The authors are members of the FORCE11 Software Citation Implementation Working Group, co-chaired by DSK.}
\IEEEauthorblockA{\textit{Institute for Software Technology} \\
\textit{German Aerospace Center (DLR)}\\
Berlin, Germany \\
https://orcid.org/0000-0003-4925-7248}
\and
\IEEEauthorblockN{Daniel S. Katz}
\IEEEauthorblockA{\textit{NCSA \& CS \& ECE \& iSchool} \\
\textit{University of Illinois}\\
Urbana, IL, USA \\
http://orcid.org/0000-0001-5934-7525}
\and
\IEEEauthorblockN{Ilian T. Todorov}
\IEEEauthorblockA{\textit{Scientific Computing Department} \\
\textit{Science and Technology Facilities Council}\\
Warrington, United Kingdom \\
https://orcid.org/0000-0001-7275-1784}
}

\author{
\IEEEauthorblockN{Stephan Druskat\IEEEauthorrefmark{1}, Daniel S. Katz\IEEEauthorrefmark{4}, and Ilian T. Todorov\IEEEauthorrefmark{3}} 
\IEEEauthorblockA{\IEEEauthorrefmark{1}Institute for Software Technology, German Aerospace Center (DLR), Berlin, Germany, ORCID: 0000-0003-4925-7248}
\IEEEauthorblockA{\IEEEauthorrefmark{4}NCSA \& CS \& ECE \& iSchool, University of Illinois, Urbana, IL, USA, ORCID: 0000-0001-5934-7525} 
\IEEEauthorblockA{\IEEEauthorrefmark{3}Scientific Computing Dept., Science and Technology Facilities Council, Warrington, UK, ORCID: 0000-0001-7275-1784}
}
\maketitle

\begin{abstract}
Software citation contributes to achieving software sustainability in two ways:
It provides an impact metric to incentivize stakeholders to make software sustainable.
It also provides references to software used in research, which can be reused and adapted to become sustainable.
While software citation faces a host of technical and social challenges, community initiatives have defined the principles of software citation and are working on implementing solutions.
\end{abstract}



\section{Introduction and context}


\noindent Software citation helps to make research software sustainable and attribute credit to its contributors.
Citations provide references to software used in, and/or created as part of, research.
Citations to research outputs are the primary impact-related metric in academia, and it is more practical to fit software into this system than to create a new one based on alternative metrics.
It is frequently assumed that the number of citations correlates positively with the impact or importance of a research output.
Because of this, citation metrics are used as a key factor in hiring, promotion and funding decisions.




Software citation has at least two effects ($E$) on research software sustainability:
\begin{enumerate*}[label=$E\arabic*$)] 
    \item \emph{Citations to software as an impact metric} incentivize staff, projects, and institutions to make software sustainable.
    \item \emph{Citations to software as references} provide identification and traceability of software used in previous research, which enables reuse and adaption.
\end{enumerate*}
Software citation also supports reproducibility:
to understand and reproduce research results, all parts of the research process including software must be uniquely identifiable, e.g., through citation.
To fully realize reproducibility, the identified software itself must also be sustainable or at least archived, so that computations can be carried out to reproduce results.


The principles of software citation are defined as
\begin{enumerate*}[label=(P\arabic*)]
    \item Importance;
    \item credit and attribution;
    \item unique identification;
    \item persistence;
    \item accessibility;
    \item specificity~\cite{smithSoftwareCitationPrinciples2016}.
\end{enumerate*}
These principles touch upon effects of software citation on both software sustainability and reproducibility. 
Concerning $E1$, (P1) is the baseline premise for research software sustainability: that software is considered an important and integral part of research that is crucial to sustain, and to fund in the future.
(P2) is required to create an attractive environment and career paths for developers and maintainers of research software. Concerning $E2$, (P3) is the baseline requirement for software identification which can lead to reuse and adaptation.
(P4) is its own aspect of software sustainability: that the unique identifier (P3) for software, and ``metadata describing the software [\ldots] should persist --- even beyond the lifespan of the software they describe.''
(P5) is required to reuse and/or adapt software. Concerning reproducibility, 
(P6) is required to facilitate access to the specific software version that was used in a research process.

\section{Software citation challenges}

\noindent Useful and usable software citation faces the following challenges. 
Software citation is not yet a standard part of the research workflow for authors\label{challenge:authors} or software developers\label{challenge:developers}.
Software also is still not always published in a way that enables citers to link to it in a citation.\label{challenge_publication}
Furthermore, the curation and quality of citation-relevant metadata cannot yet be reliably ensured, while software citation is lacking adequate support from publishers, funders, and identifier and citation infrastructures.\label{challenge:stakeholder-support}
Central to these challenges is the fact that software is not yet fully recognized as valuable research outcome, and the lack of realization that software work warrants credit through citation.\label{challenge:recognition}
There is also no consensus on what contributions and roles are sufficient for credit, especially in larger, dynamic research software projects.\label{challenge:roles}
These challenges need to be tackled at the policy, practice, and tooling levels, which is the subject of separate and future work.

The FORCE11 Software Citation Implementation Working Group is pursuing the implementation of the software citation principles, and has identified further fundamental issues for software citation, and challenges for working with stakeholders~\cite{katzSoftwareCitationImplementation2019}.
Some concern the different types and status of research software, and different means of identification for open or closed source software concepts as well as unpublished and published software. 
Others concern the scope, curation, format, and storage of software citation metadata, for all of which good working practices have not yet been found, decided upon, or established.
A challenge for establishing software citation in the scholarly culture is to address relevant stakeholders: research domains mostly need guidance; publishers need policies, metadata schemas for software citation, and guidance for their own stakeholders; repositories need solutions for mixed software/data packages, and suitable metadata schemas and infrastructure to support software deposits; indexers need type systems for identifying software as such, and a concept and metrics for counting citations with respect to software versions and overarching concepts.
Furthermore, research software that is not user-facing, such as dependencies, is often neglected in software citation, as researchers are not aware of their use of them.\label{challenge:transitive-credit}
These challenges are both social and technical, and most challenges are interdependent: social challenges can best be overcome with the help of technical solutions, and technical solutions cannot be accomplished without a strong enough social mandate, which in turn requires a change in scholarly culture in the first place.

\section{Existing solutions}

\noindent Implementations and solutions that serve some of the software citation principles exist.
(P1) and (P2) are served by ongoing arguments for the importance of software for research~\cite{anztEnvironmentSustainableResearch2020}, which find their way into policy~\cite{deutscheforschungsgemeinschaftGuidelinesSafeguardingGood2019}.
(P3) and (P6) are served by existing platforms which publish software versions with a DOI~\cite{vandesandtPracticeMeetsPrinciple2019}.
(P4) and (P5) are served by platforms and software registries that retain code, unique identifiers, and/or software metadata, and that publish software~\cite{vandesandtPracticeMeetsPrinciple2019,dicosmoSoftwareHeritageWhy2017}.

Guidance exists for specific challenges, e.g., on software citation for (paper) authors~\cite{chuehongSoftwareCitationChecklist2019} and (software) developers~\cite{chuehongSoftwareCitationChecklist2019a}. 
Software papers are a more conservative, paper-like alternative to publication of software objects~\cite{smithJournalOpenSource2018}.
Solutions to making dependencies visible and accessible for citation and credit are being developed, 
where users cite only the packages they use and those packages cite their dependencies~\cite{druskatSoftwareDependenciesResearch2020}~\cite[``Derived software'']{smithSoftwareCitationPrinciples2016}. 
Metadata schemas~\cite{jonesCodeMetaExchangeSchema2017} and formats~\cite{druskatCitationFileFormat2018}, as well as citation formats for software
have been developed. 
And the FORCE11 Software Citation Implementation Working Group continues to work with stakeholders to establish good practices, and build a social mandate for the implementation of software citation. 

\subsection{Weaknesses}

\noindent The ongoing implementation of and experience with solutions for software citation have exposed some weaknesses.
Firstly, it is hard to create consensus about solutions across disciplines and stakeholders, implement them, and promote take-up.
While the software citation needs of communities are known, the implementation of solutions is taking longer than has been hoped for.
Some particular solutions also have their own weaknesses.
For example, research software often does not have a final product that can be published;
It is developed dynamically across many versions that each can be used and each may have their own correct set of (citation) metadata.
This creates substantial overhead for developers and maintainers for the curation of software metadata.
Simultaneously, unique identifiers cannot be assigned automatically to every revision and recorded in metadata before the revision is committed.
This makes it hard to sync versions and metadata.
This can partly be solved by archiving software with Software Heritage, which assigns unique identifiers to every revision/commit.
With regard software citation metadata formats, there is no agreement which of the existing solutions~\cite{druskatCitationFileFormat2018,jonesCodeMetaExchangeSchema2017} should be used when and by which stakeholder group:
While one aims to provide only citation-relevant metadata and targets mainly human users, the other is more generic and optimized for compatibility with web standards.
This can be solved through compatibility, and existing conversion software.

\subsection{Strengths}

Many solutions for software citation address real needs.
Better software citation practice can close gaps in representing real world relations between research outputs, their impact, and the allocation of credit~\cite{druskatSoftwareDependenciesResearch2020}. 
This can lead to better career options for software developers and maintainers in academia, prevent brain drain towards industry, and thus sustain research software projects.
Many solutions are developed by and for the community. For example, 
the FORCE11 Working Group is made up of software citation stakeholders and practitioners who know the needs of their communities and/or experience them first-hand.
This will ideally lead to solutions that target the needs of communities.
Finally, software citation can show and reward the impact of software on research, and incentivize the creation and maintenance of sustainable research software, which will enable replicability and reproducibility of research results.

\bibliographystyle{IEEEtranDOI}
\bibliography{bokss-2021-software-citation.bib}

\end{document}